\documentclass[showpacs,showkeys,aps,pra,twocolumn]{revtex4-1}
\usepackage{graphicx}
\usepackage[english]{babel}
\begin{document}
\title{Testing the influence of acceleration on time dilation using a rotating  M\"{o}ssbauer absorber}
\author{ Y. Friedman}\author{J. M. Steiner} \author{S. Livshitz} \author{E. Perez}
\affiliation{Jerusalem College of Technology, P.O.B. 16031 Jerusalem 91160, Israel}
\author{I. Nowik}\author{ I. Felner}
\affiliation{Racah Institute of Physics, Hebrew University, Jerusalem 91904, Israel}
\author{ H.-C. Wille}
\affiliation{Deutsches Elektronen-Synchrotron, Notkestr. 85, D-22607 Hamburg, Germany}
\author{G. Wortmann}
\affiliation{Dep. Physik, Universit\"{a}t Paderborn, Warburger Str. 100, D-33098 Paderborn, Germany}
\author{O.Efrati} \author{ A. Finkelstein}
\affiliation{Colibri Spindles Ltd. Ind. Park Lavon,  Bikat Bet Hakerem Israel}
 \author{S. Petitgirard}
 \affiliation{Bayerisches Geo - Institut,  Universitätsstraße 30, 95447 Bayreuth, Germany}
\author{ A. I. Chumakov} \author{D. Bessas}
\affiliation{ESRF - The European Synchrotron, CS40220, 38043 Grenoble, Cedex 9, France}

\begin{abstract}

We report  here the findings  of  our recent experiment  at the Nuclear Resonance Beamline ID18 of ESRF, a sequel of our two previous experiments  at the same facility. The aim of the experiment series was to test the influence of acceleration on time dilation by measuring the relative spectral shift  between the resonance spectra of a rotating M\"{o}ssbauer absorber with acceleration  anti-parallel and parallel to the direction of the incident beam. Based on the experiences and know-how acquired in our previous experiments, 
 we collected data for rotation frequencies up to 510Hz in both directions of rotation and also used different slits. As expected, for each run with high rotation, we observed  a stable statistically significant relative shift between the spectra of the two states with opposite acceleration. This seemed to indicate the influence of acceleration on time dilation. However, we found that the observed relative shift also depends on both: the choice of the slit, as well as on the direction of rotation. These  unexpected findings, not discovered in our previous experiments, resulting from the loss of symmetry in obtaining the resonant lines in the two states, could overshadow the relative shift due to acceleration. This loss of the symmetry is caused by the deflection of the  radiative decay due to the  Nuclear Lighthouse effect from the rotating M\"{o}ssbauer absorber. We also found that it is impossible to keep the alignment (between the optical and the dynamical rotor systems) with accuracy needed for such experiment, for long runs, which resulted in the reduction of the accuracy of the observed relative shift. These prevent us to claim with certainty the influence of acceleration on time dilation using  the currently available technology. An improved KB optics with focal spot of  less than 1 micron to avoid the use of a slit and a more rigid mounting of the rotor system, are necessary for the success of such experiment. Hopefully, these findings together with the indispensable plan for a  conclusive experiment presented in the paper, will prove useful to future experimentalists wishing to pursue such an experiment.

\end{abstract}

\maketitle
\section{Introduction}

We present here the findings of  our follow-up third experiment  HC-3065 at the  Nuclear Resonance Beamline ID18 of ESRF during July 2017, aimed to test the influence of acceleration on time dilation. Should acceleration influence time dilation, it would violate Einstein's Clock Hypothesis, a cornerstone in Relativity Theory \cite{Einstein}. For experimentally feasible accelerations caused by rotation, the expected influence on time dilation is very small, so one needs very accurate methods to measure it. M\"{o}ssbauer spectroscopy is one of the best methods available  today. For technical reasons, one is forced to have the transducer outside the rotating disk with the M\"{o}ssbauer absorber on its rim. Such a setup causes the broadening of its resonant line which in turn necessitates the use of a strong  M\"{o}ssbauer source with the capability to be focused to the center of the disk. The Synchrotron M\"{o}ssbauer Source (SMS)
 \cite{SMS} together with the   KB-optics
 \cite{KB} at the  Nuclear Resonance Beamline at  ESRF 
 \cite{RC} was the best choice for our experiments.

In our previous experiments HC-1361 during July 2014  
 \cite{F14} and  HC-1898 during July 2016 
 \cite{F2016},\cite{F2017} at the  same facility, we investigated this influence  by measuring the spectral shift due to acceleration using a SMS and a specially designed rotor system incorporating a disk with a semicircular single line M\"{o}ssbauer absorber on its rim.  KB optics was used to focus the beam  to the center of the rotating disk, and a metal slit was mounted to  further narrow this beam.

In experiment HC-1361 we found 
\cite{F14} that there is a broadening of the absorption line of a rotating M\"{o}ssbauer absorber that this broadening is as predicted in 
\cite{FN}, hence in order to observe the shift one needs the entire resonant line.  It was shown that the broadening of the absorption line depends on the width of the beam at the axis of rotation. Therefore, in order to minimize this broadening, one must focus the beam to this axis  and/or to add a slit positioned along this axis, in order to block out the rays distant from this axis. We developed a methodology how to obtain the spectrum of a rotating absorber by aligning the beam with the center of rotation.

In experiment HC-1898 we improved the configuration, which in turn enabled us to reach higher rotation frequencies  and we observed 
\cite{F2017} a statistically significant  shift between the spectral lines in the \textit{two states} when the acceleration was anti-parallel and parallel to the SMS beam.
\begin{figure}[h!]
\centering
 \scalebox{0.35}{\includegraphics{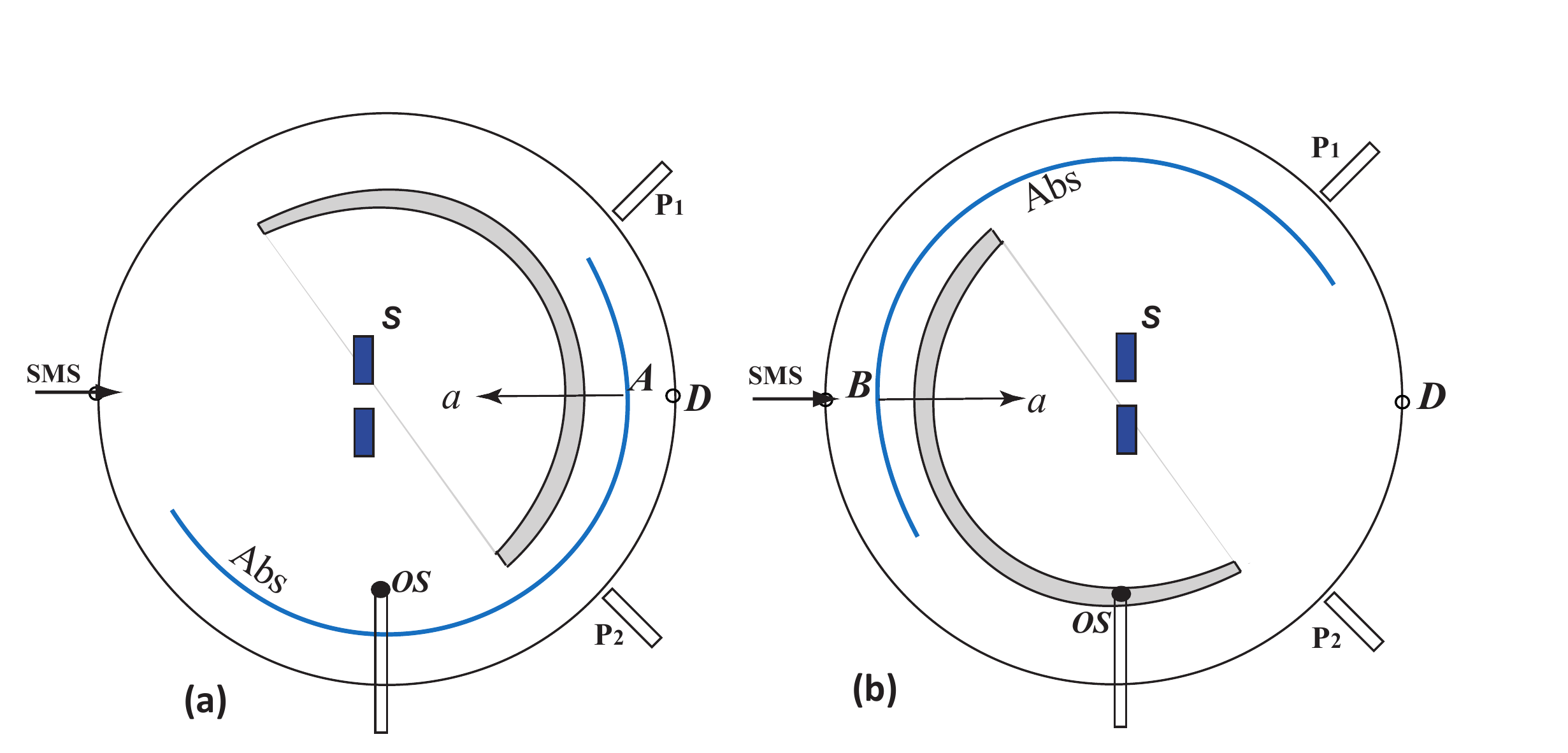}}
 \caption{The setup and two states (a) and (b), SMS source, slit $S$, semicircular absorber $Abs$, optical sensor $OS$, proximity sensors $P_1,P_2$ and detector $D$. }
 \label{TwoSt}
\end{figure}
This shift, assigned as the \textit{relative shift} (RS), remained stable for runs up to 12 hours. This result should have been enough to validate the effect of acceleration on time dilation.

 The experiment revealed however, that in fact this is not conclusive 
 \cite{F2016}. In such experiment, one must also consider the significant effect of the non-random vibrations on RS due to the imperfectness of the rotor/bearing system. We monitored these vibrations and  showed that they conform with the known forced steady state response of a Jeffcott rotor/bearing  system. Using two proximity sensors and an optical sensor we have estimated their effect on the observed RS. We found that the shift due to the vibrations of  the rotor/bearing  system was often comparable to the observed shift.  Nevertheless, in two runs of clockwise (CW) rotations at 200Hz, where the effect of vibrations was negligible, we still observed a statistically significant corrected RS of $0.41mm/s$ which is much larger than the one predicted by any model. This finding showed an indication of the influence of acceleration on time and motivated us to perform our third experiment HC-3065 with additional improvements and know-how.

Based on our observations and their implications acquired in the first two experiments, we came to the conclusion that in order to claim conclusively the influence of acceleration on time, one must fulfill every step of the following indispensable plan :
 \begin{enumerate}
   \item  obtain the full spectrum of the entire resonant absorption line of the rotating absorber;
   \item  obtain spectral lines for two separate states (a) and (b) of the rotor  (see Figure 1) differing only in the direction of their respective accelerations in order to observe the RS between them;
   \item monitor the non-random vibrations and quantify their effect on the RS;
   \item  check the reproducibility of  the RS within a similar configuration;
   \item repeat  the experiment with a configuration that should not affect the RS;
  \item  explore other effects that also lead to a RS, in case the change in configuration affects the RS;
  \item  obtain the corrected RS by eliminating any unwanted side contribution to RS;
  \item  repeat the experiment with different configurations to verify that the corrected RS is independent of the configuration;
  \item  check that the RS is unaffected by measuring a different absorber and/or rotor system;
  \item  study the dependence of the RS on the rotation frequency and disk size.
   \end{enumerate}

Steps 1,2 and 3 were addressed in our first two experiments. In experiment HC-1361 we developed a method to obtain the full resonant absorption line of a rotating absorber using SMS. In experiment HC-1898 we obtained  spectral lines for the two separate states (a) and (b), and observed the RS between them. We also discovered, that the non-random vibrations due to the imbalance of the rotor/bearing  system also cause an unwanted RS,  and found a way how to correct for it as required by step 3.

In this paper we begin with  a brief description of the improved experimental setup of experiment HC-3065, and present  additional  findings, as well as  serious unexpected results we encountered. We will show, how these relate to the later steps enumerated above, and consequently draw conclusions regarding the feasibility of a conclusive experiment. In particular we can claim with absolute confidence that there is no way to conduct a successful experiment without an improved KB optics, which will alow  to eliminate the need for a  slit. We hope this information will provide useful to future experimentalists wishing to pursue such an experiment involving a sophisticated combination of both mechanical and optical systems.

\section{The improved experimental setup}

  The basic experimental setup comprising the SMS, rotor system, vaccuum chamber, disk with a semicircular absorber, slit, KB optics and detector, was identical to that of the previous experiments with additional improvements.

 The SMS is comprised of a number of optical elements with the key element being an anti-ferromagnetic and almost ideal single crystal of iron borate $^{57}FeBO_3$, which is used in pure nuclear (111)  reflection, i.e.,  which is forbidden for electronic scattering but allowed for nuclear scattering. This crystal is maintained in an external magnetic field and is heated a bit above the Neel temperature, where the magnetic hyperfine structure collapses to a single line spectrum. The SMS at ESRF proved to be the ideal choice for such experiment as it provided a $^{57}Fe$ resonant radiation at 14.4keV within a bandwidth of 15neV. In contrast to radioactive sources, the beam emitted by the SMS is almost in full resonance and fully polarized, has high brilliance and can be focused by use of KB-optics to a $10\mu m \times 5\mu m$ spot size.

  The beam from the SMS passes the KB-optics, hits the slit and then the fast rotating absorber, when the system is in state (a), while in state (b) it hits first  the  absorber and  the slit afterwards. Finally, in each state the emerging beam is detected by the APD detector diametrically opposed to the SMS (see Figure 1).

We used the same rotor system as in the previous experiments  with a 50 mm radius disk, made of Titanium 6Al-4v which includes 108 slots, distributed uniformly around the circle, designed for rotation frequencies of up to 1kHz. A semicircular shaped enriched potassium ferrocyanide $^{57}Fe$ (95\%)  $K_4Fe(CN)_6 \cdot 3H_2O$ single- line absorber  was placed on the rim of the disk.  The disk was lowered inside a vacuum chamber, which allows 7 mbar internal pressure and is driven by a high-speed air-bearing spindle located outside the chamber. The rotor system was specially designed to control and maintain both clockwise (CW) and counterclockwise (CCW) constant rotation frequencies.

 An optical sensor allows the separation of each rotation into two states depending on the direction of the acceleration at the point of incidence of the beam with the absorber. In order to reduce broadening we  focussed the SMS beam to the centre of the disk by using the KB optics, and mounted a high-quality slit with the possibility to align its position to the axis of rotation.
The rotor system together with the vaccuum chamber were all mounted on a 6-circle Huber diffractometer. A  HUBER 5102-05 xy-stage facilitated  the mounting and adjustment of the slit with its specially designed holder, see Figure 2.

\begin{figure}[h!]
\centering
 \scalebox{0.25}{\includegraphics{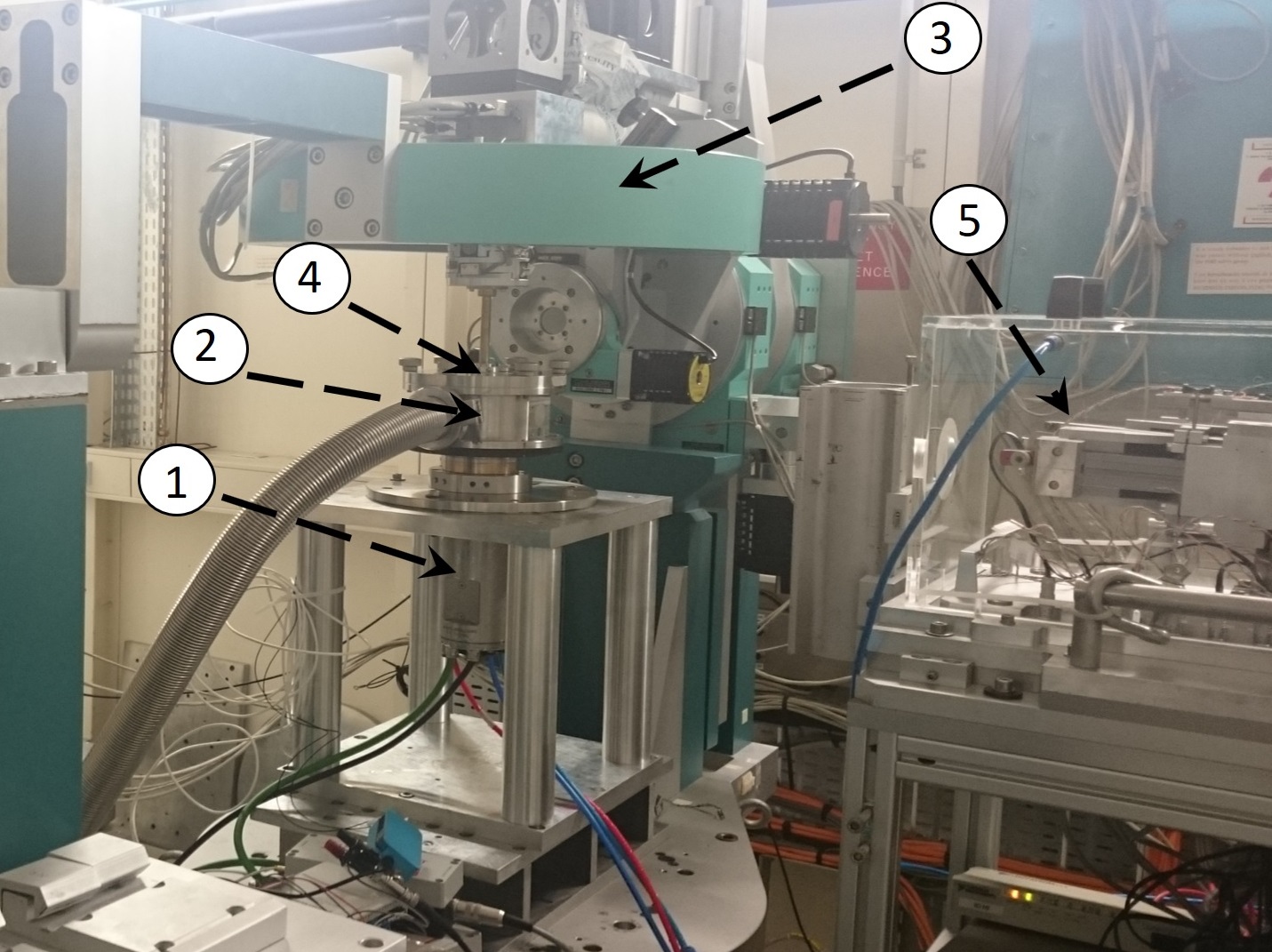}}
 \caption{The experimental system. The rotor system (1) together with the disk in vacuum chamber (2) were all mounted on a 6-circle Huber diffractometer (3). The slit (4) with its holder on top of the  vacuum chamber and  KB optics (5). }\label{Disk}
\end{figure}

A beam from the SMS  passing the KB optics crosses the disk along its diameter. As the disk rotates, the radial acceleration of the absorber  is alternately directed  anti-parallel and parallel  to the photon wavevector (see Figure 1), yielding opposite signs for the postulated additional frequency shift due to this acceleration. An optical sensor  facing  the disk mounted on the top of the chamber was used to identify the anti-parallel and parallel states (a) and (b), respectively, (see Figure 1) and the SMS data acquisition system at ESRF was modified in order to obtain simultaneously the spectra corresponding to each state. The absorption curves were measured by recording the transmitted intensity as a function of the Doppler detuning of the SMS.

The disk was driven by a high speed air bearing spindle produced by Colibri Spindles Ltd., Israel. The spindle is located outside of the vacuum chamber allowing a 7mbar internal pressure, while the driving shaft crosses the chamber wall through a carefully designed bore, leaving a gap of 50 microns between the static (chamber) and dynamic (shaft) parts. Vacuum conditions, essential for high speed rotation, low friction and acoustic noise, were maintained by an automatically controlled vacuum pump. The chamber had two openings each of width 2mm covered by transparent Mylar to allow the beam from the SMS to cross the chamber.

 In order that the slit will be adjustable,  it has to be outside the chamber. This was achieved by inserting a mini-chamber, also with two transparent openings, in the center of the chamber. The rotor system was designed to rotate in both (CW and CCW) directions by the flick of a remote switch. This also allows the annihilation of first order effects (in rotation frequency) by averaging the results produced by both directions.

Two proximity sensors (by Micro-Epsilon Ltd.) were placed orthogonal to one another in order to measure the radial displacement of the disk.  Collecting simultaneously  the data from these sensors and the optical sensor allowed us to monitor vibration at any point of the disk.

 In order to be more conclusive and get a better understanding of the RS due to rotation, we made additional improvements. The rotor system was further improved by a finer balancing of the disk and by tightening the connection between the rotor system to the supporting table. The vacuum pump in the previous experiment was replaced by a superior and stronger vacuum pump with the air pressure kept below 5.5 bar to avoid the overheating of the rotor system. We measured the vibrations of the rotor and found that they became significantly lower with these changes. The vibrations of the rotor system at different rotation frequencies were measured by using the  Schenck VIBORTEST60 tester, and we identified  the frequencies , remote from the unwanted resonant frequencies (two resonance frequency regions were identified, which are not critical, but for absolute safety and to minimize unwanted vibrations, we nevertheless worked outside these regions), for which the vibrations do not effect of the RS. All the runs were performed at these frequencies, and also enabled us to reach safely rotations up to 510Hz in both directions without  the rotor system breaking down.

 The success of the experiment requires high accuracy at high rotation frequencies. Such accuracy can only be achieved by narrowing the resonant absorption line with count rates high enough to provide statistically significant data. This narrowing for high frequencies requires a narrow slit, which in turn reduces the effective count rate. We found a way to increase significantly the count rate by modifying the SMS system, by changing the temperature of the borate crystal and by using more effective slits.

  The gold plated slit of width $10\mu m$ and thickness $15\mu m$ produced by lithography at the nano-center of the Hebrew University of Jerusalem, Israel used previously  in experiment HC-1898,  was further improved by the Intel Israel facility. This further improved the count rate, and by rotating it we were able to obtain any effective width below $10\mu m$. We also used an additional high-quality $Pt$ slit of width $2.5\mu m$ and thickness $50 \mu m$  produced by Sylvain Petitgirard at Bayerisches Geo-Institut using a focused ion beam to enable us get spectra at higher rotation frequencies and check the sensitivity of the the RS to choice of the slit.

Although, we managed to reach  sufficiently large enough counts for 510Hz,  unfortunately nevertheless, due to the limitation in the stability of the relative positioning of the rotor disk and the slit system, we observed a drift due to the loss in alignment of the beam and the rotor system for long runs. This drift caused a widening of the spectrum, which in turn reduced significantly the accuracy of the observed RS.

\section{Experimental results}

In the current experiment, steps 1,2 and 3 were addressed as in the previous experiments. The above improvements allowed us to gather more statistically significant data and high quality spectra of the rotating absorber for both states (a) and (b) for several rotation frequencies up to 510Hz in both CW and CCW directions with different slits. We collected several files for rotations of 300Hz, 360Hz and 510Hz in both directions. For example, we achieved quality spectra below ( runs 63 and 64 at 360Hz CCW and CW respectively) in both directions by using the $ 2.5 \mu m$ slit rotated by $10^\circ$ to obtain an effective slit of about $1\mu m$, see Figures 3 and 4.
\begin{figure}[h!]
\centering
 \scalebox{0.1}{\includegraphics{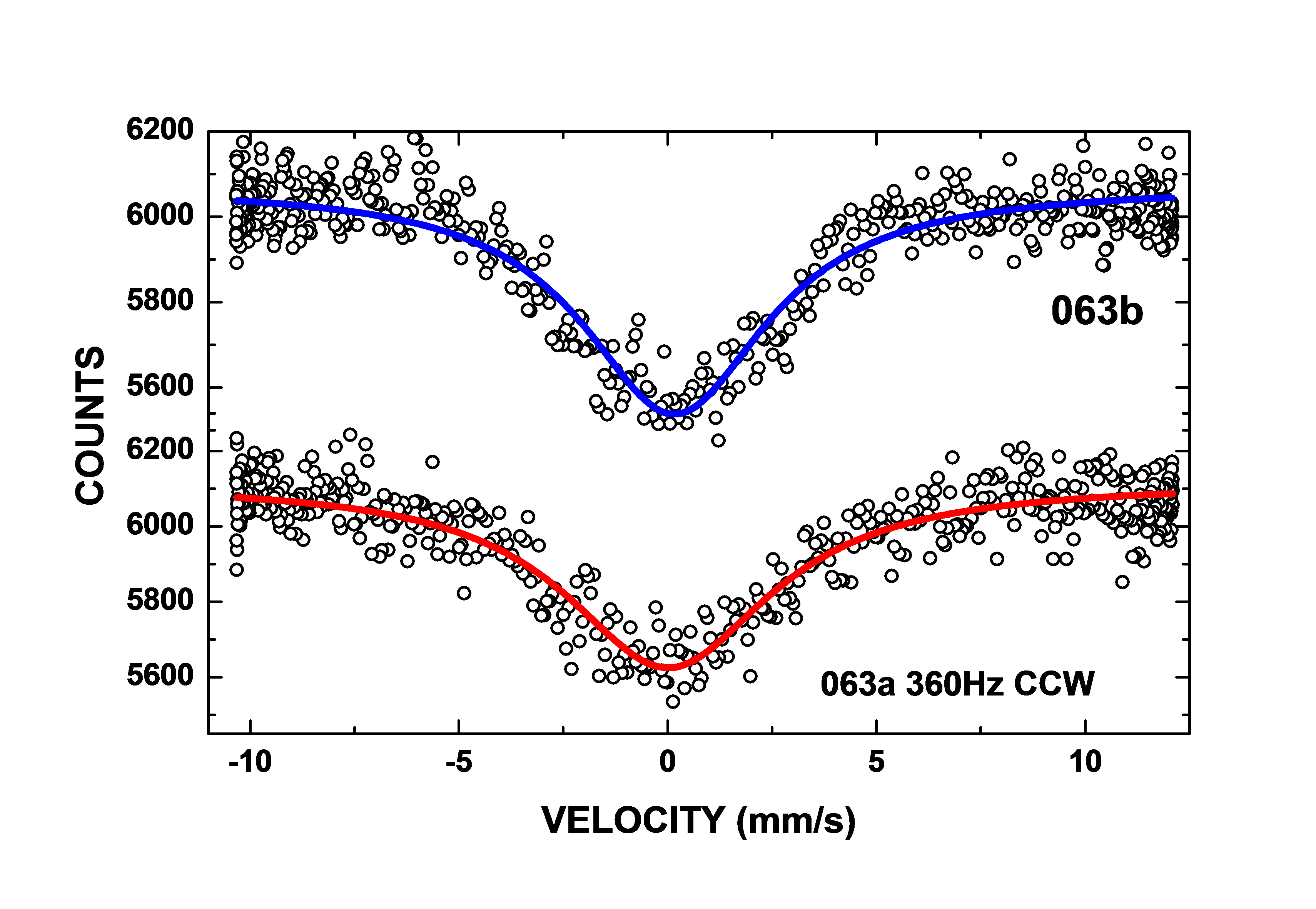}}
 \caption{The resonant lines for states (a) and (b) at 360Hz CCW rotation.}\label{33Hz}
\end{figure}
\begin{figure}[h!]
\centering
 \scalebox{0.1}{\includegraphics{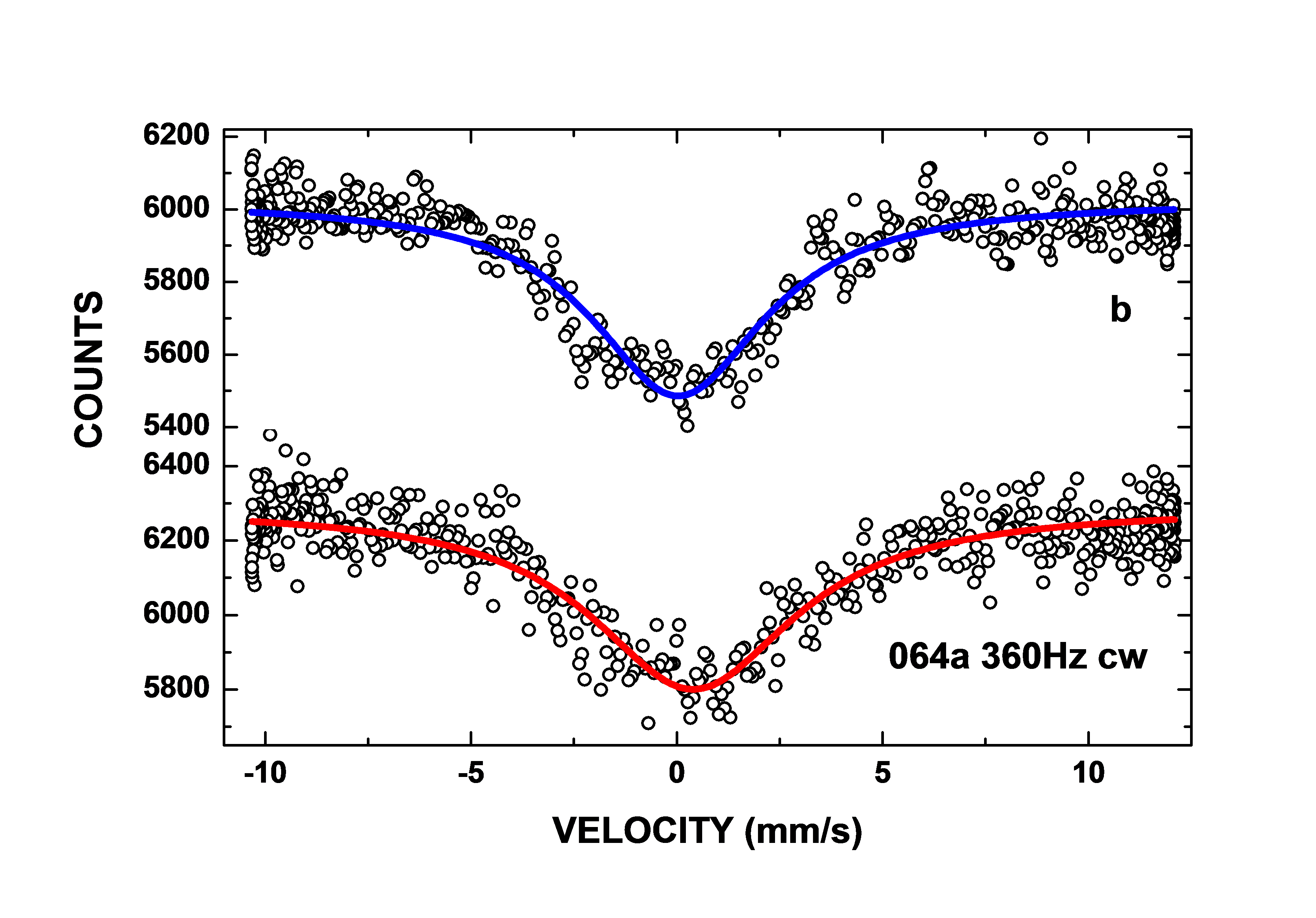}}
 \caption{The resonant lines for states (a) and (b) at 360Hz CW rotation.}\label{33Hz}
\end{figure}

 The following table summarizes the parameters of the resonant lines in the figures.
    \begin{table}[h!]
  \centering
   \caption{Parameters of resonant spectra and the vibration shifts for states (a) and (b) at 360Hz  rotation.}\label{tab3}
   \begin{tabular}{|c|c|c|c|c|}
     \hline
     Run  & Effect \%& $\gamma$ [mm/s]& Shift [mm/s]& RS[mm/s]\\
     \hline
     63a & 8.02(2) & 3.04(13) & 0.01(7) &--\\
     63b & 8.78(2) & 2.72(11) & 0.16 (6)  &-0.15(10)\\
     64a & 7.77(2) & 3.04(15)& 0.40(7) &-- \\
      64b &8.91(3) & 2.62(13)& 0.03(7) &0.37(10)\\
     \hline
\end{tabular}
\end{table}

We also checked the reproducibility of the RS suggested in step 4. We repeated the CW run at 200Hz keeping almost the same configuration  and measured exactly the same statistically significant corrected RS of $0.41mm/s$ of the previous experiment HC-1898 two years ago. Even though the system was improved, re-balanced and the gold plated  slit was improved, this corrected RS was unaltered.

As suggested in step 5, we repeated the experiment with different slits with the expectation that a change of slit should not affect the RS. The purpose of the slit  is to block out some of the rays in the beam from the KB optics remote from the  rotation axis. thereby reducing the widening of the resonant lines at higher rotational frequencies. Indeed, the use of a narrower slit reduced this widening  in all three experiments, as expected. The blocking out of the remote rays should not affect the RS for the following reason. For any ray of the beam, the first-order shift depends only on  $b$ - the distance of the ray from the rotation axis. As  shown in (Friedman \textit{et al.} 2015), the alignment shift for each ray is $\omega b$  where $\omega$ is the angular velocity vector of the rotating disk. The alignment shift of the beam is obtained by integrating this alignment shift over the distribution of  all the rays in the beam with respect to $b$. Since each (unbent) ray not blocked by the slit spends the same amount of time in each state, it has equal probability to be in either state (a) or (b). Hence,  the shift of the resonant lines in each  state (a) or (b) is expected to be identical and the RS should not be affected.

Furthermore, since the wavelength of the beam is 0.86 $\dot{A}$ and the slit width is about 1 micron, the diffraction of the beam due to the inclusion of a slit is also negligible. It is reasonable to assume that this diffraction will cause a widening of the resonant line in state (a) where the beam first hits the slit,  and the resulting diffraction increases the widening of the resonant line, see Figure \ref{TwoSt}. On the other hand in state (b), the diffraction  of the beam emerging from the absorber does not affect the resonant line, since the detector is wide enough to record even the diffracted rays.The observed slight widening of the resonant  line in state (a)  in comparison to that in state (b) (see Table 1), supports our assumption. Furthermore, it is natural to assume that the  diffraction does not brake the symmetry between the two states, therefore, the choice of slit should not affect the RS.  Even if the diffraction by the slit would brake the symmetry, changing the direction of rotation would produce the same RS with an opposite sign.

We discovered, however, that this was not the case. Without altering or tempering with any other components of the system, different slits produced different RS for the same rotation frequency. Moreover, we discovered that even in the absence of vibrations  and using the same slit, both the magnitude and sign of the RS changed with the direction of rotation.  For example,  as shown in Table 1, at 360Hz rotation the RS changed from $0.37(10)mm/s$ to $-0.15(10)mm/s$ with the change of direction from CW to CCW. This discovery revealed the existence of  a new unwanted first order RS, which cannot be ignored.

According to the currently available knowledge, a feasible explanation for this unwanted RS is based on the Nuclear Lighthouse effect of a rotating absorber discovered by Ralf R\"{o}hlsberger \textit{et al.} \cite{rohl}, \cite{rothl}.  As in the case of Nuclear Forward Scattering, the synchrotron radiation creates a collectively excited state (nuclear exciton) in the M\"{o}ssbauer isotope. In a rotating sample, these excited states acquire a phaseshift while evolving in time. The radiative decay proceeds therefore in a deflected direction. The angle of deflection depends on: 1) the rotation frequency, 2) the direction of rotation and  3) the closeness of the photon frequency to the resonant  frequency of the absorber.

In state (b) the beam from the KB-optics is  partially absorbed by the absorber and emerging beam is deflected due to the Nuclear Lighthouse effect by  angles depending on the above three factors. This deflected beam  hits the slit, which blocks out part of it before reaching the detector.  Thus, the deflection due to the Nuclear Lighthouse effect has an effect on the observed resonant  line in this state.  In state (a), however,  the beam from the KB optics is first partially blocked (and partially diffracted) by the slit  and then it undergoes a deflection after passing the absorber. This later deflection does not affect the data, since the detector is relatively large with respect to the size of the beam and is close to the absorber in this state. Hence, our assumption that  each (unbent) ray not blocked by the slit has equal probability to be in either state (a) or (b), is not valid. This fact together with the loss of symmetry in obtaining the  resonant lines in the two states, explains  the dependence of the RS on both the choice of the slit and also on the direction of rotation.

We attempted to estimate quantitatively this unwanted RS due to the Nuclear Lighthouse effect and correct for it as we did \cite{F2016}  for the RS due to non-random vibrations.  The non-random vibrations can be measured and hence the vibrational  RS could be calculated.  On the other hand, the RS due to the Nuclear Lighthouse effect depends on too many parameters of the system, which make its estimation impossible.

	As in the previous experiments, we also observed a drift in the spectrum for long runs. We discovered that this drift was substantially reduced  after tightening the screws connecting the rotor system to the plate holding it. One must have a rigid table that connects perfectly to our rotor system and preventing the screws to become loose, thereby losing the alignment of the system.

\section{Conclusions}

Experiment HC-3065 at the Nuclear Resonance Beamline ID18 of ESRF completes the series of our three experiments  at this facility.   By measuring the relative spectral shift  between the resonance spectra of a rotating M\"{o}ssbauer absorber in two states with opposite  accelerations to the  Synchrotron M\"{o}ssbauer Source, we aimed to test the influence of acceleration on time dilation. The importance and ellegance of the experiment was  pointed out \cite{RW} by Professor Rainer Weiss  of  MIT, the recipient of the 2017 Nobel prize in physics.

Using all our experiences and know-how acquired in the previous experiments  we were able to gather statistically significant data for rotation frequencies up to 510Hz in both directions of rotation without  breaking down of the system and using different slits.  For each run with high rotation, we observed  a stable statistically significant RS between the spectra of the two states with opposite acceleration. Although, this seems to indicate the influence of acceleration on time dilation as we anticipated, unfortunately, even after following a ten step systematic plan exhausting every bit of know-how acquired and the currently available technology at this facility, we cannot claim this influence conclusively. Contrary to our original assumptions, we found unexpectedly that the observed RS also depends on both the choice of the slit and on the direction of rotation. These unexpected RS, not discovered in our previous experiments, could overshadow the RS due to the acceleration, the object we were looking for. We seem to think that these RS probably result from the loss of symmetry in obtaining the  resonant lines in the two states and from the  deflection of the synchrotron beam due to the  Nuclear Lighthouse effect of a rotating M\"{o}ssbauer absorber.

  Although not fully conclusive, the experiment revealed the absolute necessity to follow every step of this indispensable plan, the impossibility of conducting this experiment using a slit to reduce the widening of the resonant lines, and the importance of maintaining the alignment between the optical and the dynamical rotor system for long runs. An improved KB optics with focal spot of 1 micron would possibly be the perfect substitute for the slit, and a more rigid mounting of the rotor system would prevent the loss of alignment and hence the spectral drift causing the widening of the spectrum thereby reducing the accuracy of the observed RS.  Furthermore, as suggested by Professor Weiss in \cite{RW}, it is important to apply active damping to reduce  the vibrations of the rotating disk  and to implement a Michelson interferometer to  measure these vibrations  accurately. Hopefully, our findings, suggested plan  and recommendations, will prove useful to future experimentalists wishing to pursue such an experiment.

  The interest in the effects of  acceleration on time dilation is ongoing. Recently, in \cite{BF} was an attempt  to interpret the observed RS in our second experiment by assuming the existence of a first order  time-dependent Doppler shift for an accelerated absorber due to the interaction time between the beam of photons and the nuclei in the absorber. This model allowed them  to calculate the  average interaction time $\tau$ of a rotating absorber from the experimentally observed relative shift RS as $\tau=RS/a$, where $a$ is the acceleration of the absorber. Alternatively, assuming that $\tau$ is affected by  the acceleration,  they suggested to calculate it from  the half-width at half-absorption intensity $\gamma$  of the  absorption resonant line of a rotating absorber. They  used  $\gamma$ of the observed  absorption line   in our experiment to calculate $\tau$, and showed that the values of $\tau$ calculated in both ways are close.

Logically, their approach is correct. However, this does not conform to results of an  experiment   performed by some members of the current  team under the leadership  by  Ralf R\"{o}hlsberger at the synchrotron PETRAIII at DESY, Hamburg in 2013. This experiment  measured the effect of acceleration on  the time spectrum (time delay) of a rotating M\"{o}ssbauer sample.  The experiment showed that  the increase of rotation frequencies does not reduce the time delay, see \cite{FIARD}. For example, for rotation frequencies  6kHz- 14kHz of a thin sample no change in the time spectrum and hence  in the average interaction time $\tau$ was observed.

The widening of the M\"{o}ssbauer absorption lines of a rotating absorber were predicted in \cite{FN}  and was verified experimentally at ESRF in \cite{F14} . The reason for this widening for a rotating absorber is the Longitudinal Doppler shift,  hence, the use of our experimental  $\gamma$  in \cite{BF}   to estimate the value of $\tau$ of a rotating absorber  is questionable.

\begin{acknowledgments}
We thank Jean-Philippe Celse from ESRF for the excellent technical support prior and during the experiment. Special thanks to Ing. Michael Livshitz, and Tal Rubinstein and Amit Alman  from the IDC Circuit Edit Lab of Intel Israel Facility for perfecting the gold plated slit used in the experiment.
\end{acknowledgments}

\end{document}